\documentclass[useAMS,usenatbib]{mn2e}
\usepackage{graphicx, setspace, subfigure, latexsym, amssymb, amsmath, booktabs, wasysym, paralist}
\usepackage{gensymb}

\title[HTRU II: Discovery of 5 Millisecond Pulsars]{The High Time Resolution Universe Survey II: Discovery of 5 Millisecond Pulsars}
\author[S. D. Bates et al.]
	{S. D. Bates$^{1}$\footnote{To whom correspondance should be addressed. Email: samuel.d.bates@student.manchester.ac.uk}, M. Bailes$^{2}$, N. D. R. Bhat$^{2}$, M. Burgay$^{3}$, S. Burke-Spolaor$^{2,4}$,\newauthor N. D'Amico$^{3}$, A. Jameson$^{2}$, S. Johnston$^{4}$, M. J. Keith$^{4}$, M. Kramer$^{1,5}$, L. Levin$^{2,4}$,\newauthor A. Lyne$^{1}$, S. Milia$^{3,6}$, A. Possenti$^{3}$, B. Stappers$^{1}$, W. van Straten$^{2}$\\
$^{1}$Jodrell Bank Centre for Astrophysics, School of Physics and Astronomy, The University of Manchester, Manchester M13 9PL, UK\\
$^{2}$Centre for Astrophysics and Supercomputing, Swinburne University of Technology, PO Box 218 Hawthorn, VIC 3122, Australia\\
$^{3}$INAF - Osservatorio Astronomico di Cagliari, Poggio dei Pini, 09012 Capoterra, Italy\\
$^{4}$Australia Telescope National Facility, CSIRO, P.O. Box 76, Epping NSW 1710, Australia\\
$^{5}$MPI fuer Radioastronomie, Auf dem Huegel 69, 53121 Bonn, Germany\\
$^{6}$Dipartimento di Fisica, Universit\`{a} degli Studi di Cagliari, Cittadella Universitaria, 09042 Monserrato (CA), Italy}
\begin{document}

\date{Accepted: 25 January 2011}

\pagerange{\pageref{firstpage}--\pageref{lastpage}} \pubyear{2010}

\maketitle

\label{firstpage}
\begin{abstract}
We present the discovery of 5 millisecond pulsars found in the mid-Galactic latitude portion of the High Time Resolution Universe (HTRU) Survey. The pulsars have rotational periods from $\sim2.3\,\rm{ms}$ to $\sim7.5\,\rm{ms}$, and all are in binary systems with orbital periods ranging from $\sim0.3$ to $\sim150$~d. In four of these systems, the most likely companion is a white dwarf, with minimum masses of $\sim 0.2\,\mathrm{M}_\odot$. The other pulsar, J1731$-$1847, has a very low mass companion and exhibits eclipses, and is thus a member of the ``black widow'' class of pulsar binaries. These eclipses have been observed in bands centred near frequencies of 700, 1400 and 3000~MHz, from which measurements have been made of the electron density in the eclipse region. These measurements have been used to examine some possible eclipse mechanisms. The eclipse and other properties of this source are used to perform a comparison with the other known eclipsing and ``black widow'' pulsars.

These new discoveries occupy a short-period and high-dispersion measure (DM) region of parameter space, which we demonstrate is a direct consequence of the high time and frequency resolution of the HTRU survey. The large implied distances to our new discoveries makes observation of their companions unlikely with both current optical telescopes and the Fermi Gamma-ray Space Telescope. The extremely circular orbits make any advance of periastron measurements highly unlikely. No relativistic Shapiro delays are obvious in any of the systems, although the low flux densities would make their detection difficult unless the orbits were fortuitously edge-on.

\end{abstract}

\begin{keywords}
pulsars: general - stars: neutron - pulsars: individual: PSR J1125--5825 - pulsars: individual: PSR J1708--3506 - pulsars: individual: PSR J1731--1847 - pulsars: individual: PSR J1801--3210 - pulsars: individual: PSR J1811--2405
\end{keywords}

\section{Introduction}
Millisecond pulsars (MSPs) are neutron stars (NSs) with rapid rotation rates that are believed to be formed in binary systems when the NS accretes matter from the companion, causing a `spin-up' in the NS's rotation rate \citep[e.g.\,][]{alpar1982}. This process results in pulsars with spin periods between 1 and 100~ms and magnetic field strengths less than $10^{10}$~G. Those systems with spin periods less than about 20~ms are thought to have had low mass companions, while those with periods between 20 and 100~ms are thought to have been spun-up by a heavy white dwarf (WD) or NS companion. During the spin-up phase, the heating caused by accretion leads to the emission of X-ray radiation \citep{do1973}. These accreting systems are known as high-mass and low-mass X-ray binaries depending upon the companion mass (HMXBs and LMXBs respectively, see \citet{bvdh1991} for details of evolution). The link between LMXBs and MSPs was not confirmed until pulsations with a period of 2.4~ms were observed in the accreting X-ray binary, SAX~J1808.4$-$3658 \citep{wvdk1998}. This link has more recently been reinforced by the discovery of PSR~J1023$+$0038 \citep{asr2009}, where radio emission is likely to have only switched on recently after an LMXB phase.

In contrast to this evolutionary model, however, around 20\% of the MSP population appear to be isolated bodies, including the first MSP to be discovered, PSR~B1937+21 \citep{backer1982}. Therefore, if this scenario is correct, and the isolated MSPs
are the descendants of the eclipsing MSPs, we might expect some of them to eventually ablate their companions until nothing remains \citep{rst1989}. The discovery of the so-called `black widow' pulsar, PSR~B1957$+$20 \citep{fst1988} and other similar systems \citep[e.g.\,PSR~J2051$-$0827,\,][]{bws1996}, provides some evidence of this process taking place.

In these systems, the pulsar's companion is typically of very low mass ($\sim 0.02\,\mathrm{M}_\odot$), and the pulsar is eclipsed by the companion for at least 10\% of the orbit. As the pulsar approaches eclipse, the pulses experience a delay due to the additional ionised gas through which the radiation has to pass, while outflowing material from the companion often gives rise to anomalous eclipses away from superior conjunction. However, the timescale over which the ablation process would lead to the companion being destroyed is far too long to explain the abundance of isolated MSPs \citep{bws1996}. Clearly other mechanisms such as tidal disruption are needed to explain the existence of this group.

The discovery of PSR~J1903$+$0327 \citep{champion2008} further challenges the conventional formation scenarios. Not only is the pulsar's companion a main-sequence star, but the orbit is also highly eccentric ($e=0.44$), neither of which are predicted by the standard formation scenarios. Three alternative theories for the formation have been proposed;
\begin{inparaenum}[\itshape a\upshape)]
\item the pulsar was born in an eccentric orbit, spinning rapidly;
\item the pulsar was spun-up in a globular cluster before being ejected into the Galactic disk; and
\item the recycling of the pulsar in a triple system \citep{champion2008}.
\end{inparaenum}
\citet{freire2010} have since confirmed that the main sequence star is the companion, and excluded the possibility that PSR~J1903+0327 is currently a member of a triple system, but maintain that the pulsar was born in a triple system, from which the component originally responsible for the spun-up of PSR~J1903+0327 has been ejected. However, if the alternatives are viable, we need to understand how many MSPs form via this channel and to do that a larger sample of MSPs is required.

Beyond improving our understanding of binary star evolution, there are many areas of pulsar science which benefit from the discovery of recycled pulsars. The discovery of binaries with, for example, a neutron star or black hole companion offer the chance to test General Relativity (GR) and other theories of gravity in the strong field regime \citep[e.g.\,][]{kramer2004,kramer2006}. The case of a pulsar with a WD companion also offers the possibility of observing effects predicted by GR. For example, PSR~J1909$-$3744 is in a 1.5~day orbit with a companion of mass $0.2\,\mathrm{M}_\odot$, where Shapiro delay has been observed, allowing the measurement of both the companion and pulsar masses \citep{jacoby2003, jacoby2005}.

The NS equation of state is currently poorly understood, and the predictions of NS masses and radii vary dramatically for different models \citep{lp2004}. One way to place a limit on these parameters is to identify the limiting rotation frequency, beyond which the neutron star would break apart. Currently, the most rapidly-rotating known pulsar is PSR J1748$-$2446ad \citep{hrs+06}, which rotates at 716~Hz; the discovery of an MSP rotating even faster than this could rule out some NS equation of state models \citep{lp2007, lp2010}. By constraining the maximum mass of a NS, it should also be possible to eliminate many models of the equation of state \citep{lp2004}. To do this, one can use binary effects such as Shapiro delay which allow the pulsar mass to be measured \citep[e.g.\ ][]{lyne2004}; an effect which is most measurable for those systems where the pulsar is found to have a massive companion or a near edge-on orbit \citep[e.g.\,][]{demorest2010}. Therefore, the discovery of more systems where the NS mass can be measured \citep[e.g.\,][]{freire2010} would contribute to the understanding of the NS equation of state. For some pulsars in binary systems, it has been possible to observe the companion optically \citep[e.g.\,][]{bassa2006} and measure its radial velocity as a function of orbital phase. However, for WD companions, often the faintness of the companion only allows the measurement of a temperature, from which a mass must be estimated using evolutionary models \citep{vankerkwijk1996}.

Currently, pulsar timing arrays \citep{hobbs2009,ferdman2010,jenet2009} are attempting to detect gravitational waves by the correlation of arrival times \citep[e.g.\,][]{jenet2006} from pulsars. These arrays require long-term, high precision timing observations with high signal-to-noise ratio; MSPs, with their short spin periods and stable rotation rates, are well-suited for these timing arrays \citep{verbiest2009}. However, not all MSPs have the narrow pulse profiles, regular rotation and high flux density required to obtain high timing precision, nor is the distribution of the known sources as uniform over the sky as desired \citep[see, e.g.\,][]{hd1983}. Hence, further discoveries from pulsar surveys may make important contributions to pulsar timing arrays and improve their sensitivity to the stochastic background of gravitational waves.

The largest previous pulsar survey, that found many of the known MSPs, was the Parkes Multibeam pulsar survey (PMPS) by \citet{mlc+01}, which surveyed the area of the Galactic plane bounded by $260\degree \leq l \leq 50\degree, |b| \leq 5\degree$, discovering 26 MSPs. At higher Galactic latitudes, the Swinburne Intermediate-latitude pulsar survey ($260\degree \leq l \leq 50\degree, 5\degree \leq |b| \leq 15\degree$), using a similar observing system except for a reduced sampling time, discovered a further 8 MSPs \citep{ebsb01,eb2001}. However, these surveys were limited by the relatively broad frequency channels (3~MHz) and coarse time sampling (250~$\mu$s and 125~$\mu$s, respectively) compared with today's digital backends, which provide increased sensitivity to distant and rapidly-rotating pulsars. 

The ionised interstellar medium disperses the pulses emitted by radio pulsars as they traverse it. When removing this dispersion, increased frequency resolution allows this correction to be made with reduced smearing of the pulses. The increased time resolution provides greater sensitivity to short-period pulses, such as those from MSPs, and by sampling with 2 bits (compared to 1 bit per sample in the PMPS) sensitivity is further increased by reducing losses due to digitisation. The High Time Resolution Universe Survey \citep[HTRU, ][]{keith2010} aims to make use of these improvements in backend technology to perform a survey of the entire southern sky with much-improved sensitivity to MSPs.

The ongoing PALFA survey at Arecibo \citep{cordes2006} and the GBT 350-MHz survey \citep{boyles2010} have each discovered several MSPs which occupy the short-period and high-DM region of parameter space which we look to probe with HTRU. In particular, the discovery, as previously mentioned, of PSR~J1903$+$0327 \citep{champion2008} as part of PALFA stands out with a rotation period of 2.15~ms and a DM of $297.5\,\mathrm{cm}^{-3}\,\mathrm{pc}$ as evidence that time resolution of 64~$\mu$s allows this parameter space to be probed.

In this paper we outline the spin and orbital parameters of five MSPs discovered as part of the HTRU survey, compare the properties of these pulsars to the previously-known population, and study in detail the eclipses that one of them displays. These discoveries have been made with $\sim30\%$ of the survey region observed, indicating that we might expect tens of MSPs to be discovered when the survey is completed.

\begin{table}
	\begin{center}
	\caption{Observational parameters for the mid-latitude portion of the HTRU survey.}
		\begin{tabular}{lr}
		\toprule
		Number of beams  & 13 \\
		Polarizations/beam   & 2 \\
		Centre Frequency & 1352 MHz\\
		Frequency channels  & 1024 $\times$ 390.625 kHz* \\
		\midrule
		Galactic longitude range & $-120\degree$ to $30\degree$ \\
		Galactic latitude range & $|b| \leq 15\degree$ \\
		Sampling interval & 64 $\mathrm{\mu}$s \\
		Bits/sample & 2 \\
		Observation time/pointing & 540 s \\
		\bottomrule
		\end{tabular}
		\label{table:survey}
	\end{center}
	*154 of these channels are then masked to remove interference
\end{table}

\section{Observations}
\subsection{Discovery and Timing}
The HTRU survey \citep{keith2010} is broken into three distinct components with different scientific objectives. The low latitude region covering Galactic latitude $|b|<3.5\degree$, the mid-latitude region covering $|b|<15\degree$, and the high latitude region which covers the remaining sky below declination $+10\degree$. The mid-latitude survey is designed to find millisecond pulsars that were missed by previous surveys due to excessive dispersion smearing but are still bright enough to be found in $\sim 10\,\mathrm{min}$. Observations are made using the 13-beam multibeam receiver on the 64-metre Parkes radio telescope, which has a half-power beam width of $\sim 0.23\degree$. Data were then processed using the \textsc{hitrun} pipeline described in \citeauthor{keith2010}; all five MSPs were discovered in the mid-latitude component of this survey, which covers an area of the Galactic plane bounded by $-120\degree < l < 30\degree, |b| \leq 15\degree$ with observations 540~s in duration. Full details of the survey parameters are given in Table~\ref{table:survey}.

After their discovery and confirmation, timing observations of these MSPs were made with baselines ranging from 250 days to more than 400 days. Those which are visible from Jodrell Bank Observatory --- PSRs J1731$-$1847, J1801$-$3210 and J1811$-$2405 --- were regularly observed with the 76-metre Lovell Telescope in a band centred at 1524~MHz, while PSRs J1125$-$5825 and J1708$-$3506 were observed with the 64-metre Parkes radio telescope in a band centred at 1369~MHz (these Jodrell Bank and Parkes data are, hereafter, `the 20~cm band'). At Jodrell Bank, observations were made approximately once per week, and Parkes observations were more sporadic, with gaps between observations sometimes lasting as long as three months. Timing observations at both observatories were made using digital filterbanks (DFBs). Flux-calibrated observations using the Parkes DFB were used for measurement of the source fluxes. Calibration was performed using routines included in \textsc{psrchive} \citep{psrchive2004} and with observations of both a calibrator source (Hydra A) and a pulsed cal.

In order to study the pulse profile evolution with frequency, observations were made of some of the sources using the 10/50~cm band receiver mounted on the Parkes telescope (the bands have centre frequencies of 732 and 3094~MHz), and using the Westerbork Synthesis Radio Telescope (WSRT) and the PuMA II backend \citep{karup2008} in a band centred at 347~MHz (hereafter, `the 92~cm band'). Observations were made with the WSRT for each of the pulsars with $\delta>-35\degree$, however, neither PSR J1801$-$3210 nor PSR J1731$-$1847 have been detected at this wavelength. The system parameters for all the observations are shown in Table~\ref{table:timing}.

The timing solutions for each of the systems are shown in Table~\ref{fullsolns}. The first quantities given in this table are the positions of the sources after timing over the specified range of dates. This timing has allowed the positions to be fitted to sub-arcsecond precision, with positional errors as given. Four of these new discoveries lie within $5\degree$ of the Galactic plane, with only PSR~J1731$-$1847 outside this range, with $b=8.15\degree$.

\begin{table}
	\begin{center}
	\caption{Observing system details for the timing observations made as part of this work. Note the specifications for the Lovell Telescope take into account the removal, as standard, of a section of the observing bandwidth.}
		\begin{tabular}{lcccc}
		\toprule
		Telescope  & Centre Freq. & BW & N$_\mathrm{chans}$ & $\langle t_\mathrm{obs} \rangle$ \\
		&(MHz) &(MHz) & &(s)\\
		\midrule
		Parkes 64-metre & 732 & 64 & 512 & 900 \\
		& 1369 & 256 & 1024 & 600 \\
		& 3094 & 1024 & 1024 & 900\\
		\midrule
		Lovell Telescope & 1524 & 384 & 768 & 900 \\
		\midrule
		WSRT & 345.625 & 80* & 512 & 2400 \\
		\bottomrule
		\end{tabular}
		\label{table:timing}
	\end{center}
	* With coherent dedispersion
\end{table}

\begin{table*}
	\caption{Observed and derived parameters for the new millisecond pulsars. The DM distance has been estimated using the model of \citet{ne2001}, while a pulsar mass of $1.4\,\mathrm{M}_\odot$ has been assumed in calculating companion masses.}
		\begin{tabular}{lccccc}
		\toprule
		Parameter & J1125$-$5825 & J1708$-$3506 & J1731$-$1847 & J1801$-$3210 & J1811$-$2405\\
		\midrule
		Right Ascension (J2000) & 11:25:44.3654(3) & 17:08:17.623(1) & 17:31:17.6072(2) & 18:01:25.8890(2) & 18:11:19.8539(2) \\
		Declination (J2000) & $-$58:25:16.867(5) & $-$35:06:22.85(7) & $-$18:47:32.74(1) & $-$32:10:53.72(1) & $-$24:05:18.72(7) \\
		Galactic Longitude ($\degree$) & 291.89 & 350.47 & 6.89 & 358.92 & 7.07 \\
		Galactic Latitude ($\degree$) & 2.60 & 3.12 & 8.15 & $-$4.58 & $-2.56$\\
		\\
		Discovery S/N & 11 & 18 & 14 & 11 & 13 \\
		Offset from survey& $0.12$ & $0.05$ & $0.06$ & $0.03$ & $0.17$\\
		beam centre ($\degree$) & & & & & \\
		\\
		TOA Range (MJD) & 55135--55431 & 55127--55461 & 55148--55399 &54996--55409 & 55131--55389 \\
		\\
		$P$ (ms)& 3.1022139189504(8) & 4.50515894826(6) & 2.3445595833757(6) & 7.45358437341(2) & 2.66059331690(1) \\
		$\dot{P}$ ($\times 10^{-20}$) & 5.963(5) & 2.3(4) & 2.49(1) & 0.265(97) & 1.36(2) \\
		DM ($\rm{cm}^{-3}\,\rm{pc}$) & 124.81(5) & 146.8(2) & 106.56(6) & 176.7(4) & 60.64(6)\\
		\\
		DM Distance (kpc) & 2.6 & 2.8 & 2.5 & 4.0 & 1.8 \\
		$\rm{S}_{1.4\,\mathrm{GHz}}$ (mJy) & 0.44(2) & 0.33(1) & 0.60(2) & 0.21(1) & 0.37(1)	 \\
		$\rm{L}_{1.4\,\mathrm{GHz}}$ (mJy kpc$^2$) & 3.0 & 2.6 & 3.8 & 3.4 & 1.2 \\
		\\
		$\tau_\mathrm{c}$ (yr) & $8.2\times 10^8$ & $3.1\times 10^9$ & $1.5\times 10^9$ & $4.5 \times 10^{10}$ & $3.1 \times 10^9$ \\
		$B_\mathrm{surf}$ (G) & $4.3\times 10^{8}$ & $3.2 \times 10^{8}$ & $2.3 \times 10^{8}$ & $1.4 \times 10^{8}$ & $1.9 \times 10^8$ \\
		$B_\mathrm{lc}$ (G) & $1.3 \times 10^5$ & $3.2 \times 10^4$ & $1.7 \times 10^5$ & $3.1 \times 10^3$ & $9.3 \times 10^4$\\
		$\dot{E}$ (erg s$^{-1}$) & $7.9\times 10^{34}$ & $9.9\times 10^{33}$ & $7.6\times 10^{34}$ & $2.5\times 10^{32}$ & $2.8\times 10^{34}$ \\
		$\dot{E}$/$\rm{d}^2$ (erg kpc$^{-2}$ s$^{-1}$) & $1.2\times 10^{34}$ & $1.2\times 10^{33}$ & $1.2\times 10^{34}$ & $1.6\times 10^{31}$ & $8.6\times 10^{33}$ \\
		\\
		Binary Model & ELL1 & ELL1 & ELL1 & ELL1 & ELL1 \\
		\\
		Orbital Period (d) & 76.4032169(4) & 149.13318(1) & 0.311134124(4) & 20.7716995(3) & 6.27230204(2)\\
		$a\sin{i}$ (lt-s) & 33.638356(1) & 33.584233(5) & 0.1201594(9) & 7.809320(5) & 5.7056623(8)\\
		TASC (MJD) & 55126.348808(1) & 55132.23096(2) & 55132.3113000(8) & 55001.934481(3) & 55136.1686237(1) \\
		$\epsilon_1$ & $-0.00025349$(5) & 0.0000004(2) & 0.00010(1) & $-0.000003$(1) & 0.0000017(4)\\
		$\epsilon_2$ & $-0.00004404$(6) & $-0.0002445(2)$ & 0.00000(1) & 0.000000(1) & 0.0000011(4)\\
		$e$ & 0.00025720(5) & 0.0002445(2) & 0.00014(1) & 0.000004(1) & 0.0000024(4) \\
		$\omega$ (deg) & 260.140(2) & 179.91(5) & 264(5) & 290(20) & 253(7) \\
		\\
		Min. $m_\mathrm{c}$ ($\mathrm{M}_\odot$) & 0.26 & 0.16 & 0.037 & 0.15 & 0.24 \\
		Med. $m_\mathrm{c}$ ($\mathrm{M}_\odot$) & 0.30& 0.19& 0.043& 0.17&  0.28\\\\
		RMS of fit ($\mu$s) & 5.538 & 7.771 & 6.339 & 29.050 & 5.539 \\
		$\chi^2$ of fit & 0.95 & 1.3 & 7.8 & 1.3 & 1.3 \\
		\bottomrule
		\end{tabular}
		\label{fullsolns}
\end{table*}

\subsection{Orbital and Spin Parameters}
\begin{figure}
	\begin{center}
	\includegraphics[width=8cm]{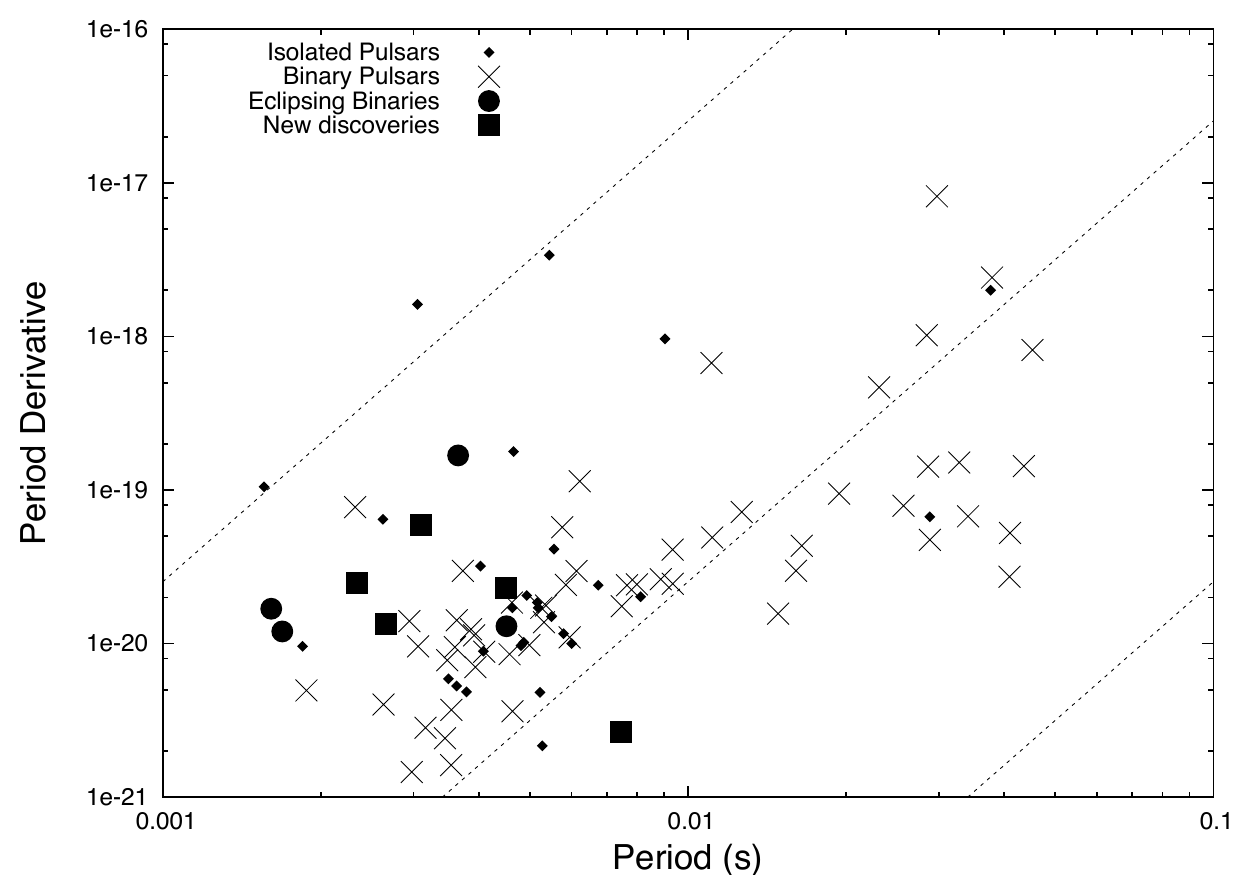}
	\end{center}
	\caption{$P$--$\dot{P}$ diagram of the millisecond pulsars, divided by binary `type'. Lines of constant $\dot{E}$ cross the population at (from left to right) $10^{36}$, $10^{33}$ and $10^{30}\,\rm{erg}\,\rm{s}^{-1}$. Pulsars in globular clusters have been excluded from this figure.}
	\label{fig:ppdot}
\end{figure}

The pulse periods of the newly discovered pulsars range from 2.3~ms (PSR J1731$-$1847) to 7.5~ms (PSR J1801$-$3210) (see Table~\ref{fullsolns}). The change in period seen in confirmation observations of all five sources made it apparent that the pulsars are in binary systems. With subsequent observations, we were able to fit for the observed variation in pulse period with time and deduce a set of orbital parameters for each pulsar. In order to precisely measure the spin and orbital parameters of each pulsar, the regular timing observations in the 20~cm band were used to produce a set of time-of-arrival (TOA) measurements. Using the standard pulsar timing procedure \citep[e.g.\ ][]{lk05}, these TOAs were used to fit for each of the parameters presented in Table~\ref{fullsolns} using the Tempo2 pulsar timing package \citep{tempo2_1}. The locations of these pulsars in the $P$--$\dot{P}$ plane are shown in Fig.~\ref{fig:ppdot}, and are seen to lie towards the lower end of the period distribution, with each of the new discoveries fitting in the region traced by the previously-known population with short spin periods.

The transverse velocity, $V_\mathrm{T}$, of a pulsar at distance $d$ makes a contribution to the measured value of $\dot{P}$ in what is known as the Shklovskii effect \citep{shklovsky1970}
\begin{equation}
\frac{\dot{P}}{P} = \frac{1}{c}\frac{V_\mathrm{T}^2}{d}.
\label{shklovsky}
\end{equation}
We can then calculate the value of $V_\mathrm{T}$, given some contribution to $\dot{P}/P$. Assuming this contribution is 10\% for each MSP, we find that PSRs~J1708$-$3506 and J1731$-$1847 would be required to have very large values of $V_\mathrm{T}$, 670 and 350~$\mathrm{km\,s}^{-1}$ respectively. The other three MSPs would require $V_\mathrm{T}<100\,\mathrm{km\,s}^{-1}$, which are `realistic' values \citep{toscano1999}; and if the Shklovskii contribution was as high as 50\% for these three, $V_\mathrm{T}$ would only need be $\sim 100\,\mathrm{km\,s}^{-1}$. It seems feasible, therefore, that there is a significant Shklovskii contribution to the measured value of $\dot{P}$ for these pulsars, which in turn implies that the values of $\dot{E} \propto \dot{P}$ and $B_\mathrm{surf} \propto \dot{P}^{1/2}$ are overestimated and $\tau_\mathrm{c} \propto \dot{P}^{-1}$ underestimated in Table~\ref{fullsolns}.

In this work we have used the ELL1 pulsar timing model (as outlined in the appendix of \citealt{lange2001}) to fit for all the relevant spin and binary parameters in each of the pulsars. This model is suitable for modelling the orbits of low-eccentricity systems, and for PSRs J1125$-$5825 and J1708$-$3506, gives a statistically significant value for the orbital eccentricity. With the exception of PSR J1731$-$1847, the reduced $\chi^2 \sim 1$ for each of the timing models. In the case of PSR J1731$-$1847, the higher $\chi^2$ can be attributed to the extra delay induced by the eclipses (as discussed below). With further observations, and taking care to remove TOAs around the eclipse region, the $\chi^2$ value is expected to reduce.

The orbital parameters for the five MSPs are shown in Table~\ref{fullsolns}, from which it can be seen that they have a wide range of orbital periods, from 0.3~d (PSR~J1731$-$1847) to 150~d (PSR~1708$-$3506). Using the orbital parameters for each pulsar, we can relate the mass of the pulsar, $m_\mathrm{p}$, and the mass of its companion, $m_\mathrm{c}$, to the projected semi-major axis of the orbit, $x$, using the Keplerian mass function 
\begin{equation}
f(m_\mathrm{p}, m_\mathrm{c}) = \frac{(m_\mathrm{c} \sin i)^3}{(m_\mathrm{p} + m_\mathrm{c})^2} = \frac{4 \pi^2}{G} \frac{x^3}{P_\mathrm{b}^2}
\label{eq:massfunction}
\end{equation}
where G is Newton's gravitational constant and $P_\mathrm{b}$ is the orbital period of the pulsar in the binary system, and $i$ the inclination of the orbit. Assuming a pulsar mass of 1.4~M$_\odot$, the mass function may be solved for the mass of the companion as a function of orbital inclination. Solving for $i=90\degree$ gives the minimum mass of the companion, and $i=60\degree$ gives the median mass; each of these values are given in Table~\ref{fullsolns}. Apart from PSR J1731$-$1847, the companions to these pulsars have median mass values of $\sim 0.2\,\mathrm{M}_\odot$, indicating that they are typical pulsar-WD binary systems. PSR J1731$-$1847, however, has a companion with a very low median mass, only $0.043\,\mathrm{M}_\odot$. As this low mass suggests \citep[e.g.\,][]{rt1991, bws1996}, PSR J1731$-$1847 displays eclipses, which are discussed in detail in section~\ref{sec:1731}.


\begin{figure*}
	\begin{center}
	\includegraphics[width=17cm]{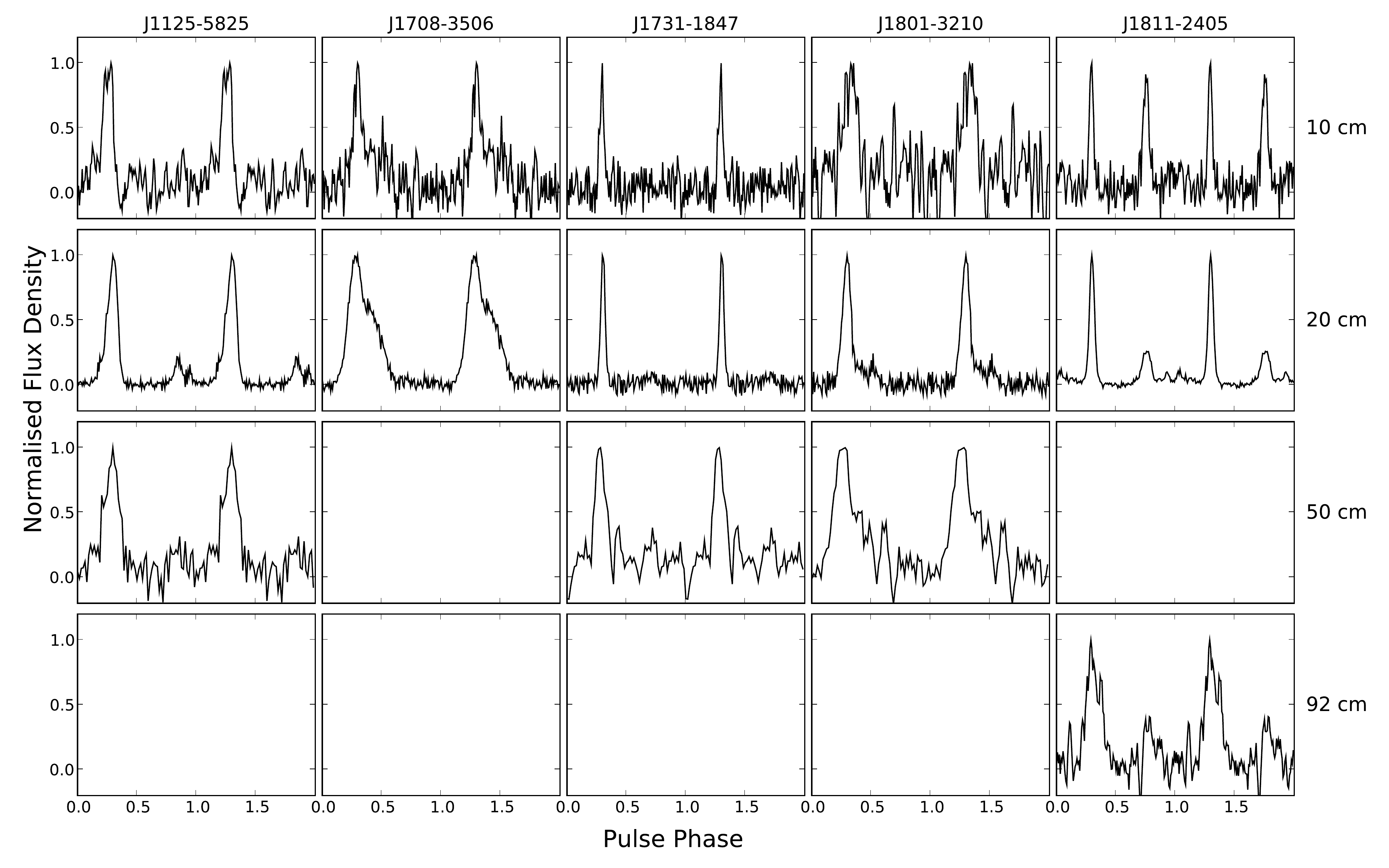}
	\end{center}
	\caption{Average pulse profiles, where available, for each of the new MSPs from observations at wavelengths of 10~cm, 20~cm and 50~cm using the Parkes radio telescope, and at 92~cm using the Westerbork Synthesis Radio Telescope. In each case, the profile is not flux calibrated, and the peak has been placed at phase 0.3. No observations have been made at 50~cm for PSRs~J1708$-$3506 and J1811$-$2405.}
	\label{profiles}
\end{figure*}

\subsection{Pulse Profiles}\label{sec:pulseprofs}
The pulse profiles of each pulsar in the different observing bands are shown in Fig.~\ref{profiles}. PSRs J1125$-$5825 and J1811$-$2405 each display an interpulse, trailing the main pulse by $\sim0.55$ and $\sim0.45$ in pulse phase, respectively. The spectral properties of the two interpulses is different at shorter wavelengths --- the interpulse of PSR J1125$-$5825 is not visible in the 10~cm band, probably due to the low S/N of the profile. However, in PSR J1811$-$2405 the flux of the interpulse increases relative to the main pulse at 10~cm, indicating that the interpulse has a flatter spectrum than the main pulse. This has been observed previously, both for normal pulsars \citep[e.g.\,][]{biggs1988} and MSPs J2322$+$2057 \citep{nice1993}, J1012$+$5307 \citep{kramer1999} and B1855$+$09 \citep{kijak1997}. The interpulse of PSR J1125$-$5825 also appears to be complex, consisting of at least 2 components in the 20~cm band, while the profile of PSR J1811$-$2405 displays off-pulse emission trailing the interpulse that can be seen in the pulse profiles at 10 and 20~cm.

The main peaks of the pulse profiles of PSRs J1731$-$1847 and J1811$-$2405 are narrow in comparison to the other new discoveries. This property can make high-precision timing of an MSP an easier task, however the eclipses of PSR J1731$-$1847 make it unlikely to be suitable for pulsar timing arrays. The profiles of PSRs J1708$-$3506, at 20~cm, J1801$-$3210, at 50~cm, and J1811$-$2405, at 92~cm, display broadening of the pulse with respect to the profiles at shorter wavelengths. In these cases, this broadening can be attributed to the scattering of pulses by the interstellar medium, as the high frequency resolution of the observing systems precludes the possibility that this is caused by residual DM smearing across a frequency channel, while the observation in the 92~cm band has been coherently dedispersed.

Using the empirical fit of \citet{bcc+04}, we would expect the scattering tail in these observations to have a $1/e$ timescale of 0.7~ms for PSR J1801$-$3210 and 0.1~ms for PSR J1811$-$2405, but only 0.02~ms for PSR~J1708$-$3506. Since there is a large amount of scatter around the Bhat et al.\,relationship, the fact that this timescale is slightly underestimated in the first two cases each case is not likely to be significant --- for comparison, the NE2001 electron distribution model \citep{ne2001} predicts scatter broadening times of 0.01, 0.4 and 0.5~ms for PSRs J1708$-$3506, J1801$-$3210 and J1811$-$2405 respectively. It is also possible that this pulse broadening is due to intrinsic evolution of the pulse profile with observing frequency, and for PSR~J1708$-$3506, it is unclear whether this broadening is due to a second component which trails the main peak. This component may be present, though faint, in the 10~cm observation. 

For the two pulsars which were observed, but not detected, at 92~cm, assuming the standard relationship for pulse scattering as a function of wavelength \citep[e.g.\,][]{lr+2000}, $\tau_\mathrm{s} \propto \lambda^{(11/3)}$, the $1/e$ timescale for scattering, $\tau_\mathrm{s}$, should be increased by a factor of $\sim 11$ at 92~cm compared to 50~cm. In the case of PSR J1801$-$3210, it seems likely that the increased scattering could be responsible for smearing the pulse over more than a pulse period. For PSR J1731$-$1847, however, the observations were made well away from orbital phase 0.25 (removing the possibility of eclipse), nor does it appear that the scattering would be severe enough to smear the pulse by more than a pulse period, making it likely that the flux density at this wavelength was below the detection threshold, implying a relatively flat spectral index, or a spectral break.

\subsection{Potential measurement of relativistic effects}\label{sec:GReffects}
In certain cases, with favourable orbital parameters, it is possible to measure the mass of both the pulsar and its companion \citep[e.g.\,][]{demorest2010}. When light passes near a massive body, the path that the light must travel is increased in length, due to the curvature of space-time caused by the massive body. This effect is known as Shapiro delay, and may be measured for near edge-on binary systems containing an MSP. For low-eccentricity systems, the Shapiro delay, $\Delta_\mathrm{SB}$, at orbital phase $\Phi$ is \citep[e.g.\,][]{taylor1992}
\begin{equation}
\Delta_\mathrm{SB} = -2r \ln [1-s\sin\Phi]
\label{eq:shapiro}
\end{equation}
where $r$ and $s$ are post-Keplerian orbital parameters,
\begin{equation}
r = \frac{G m_\mathrm{c}}{c^3}\mathrm{M}_\odot,
\label{eq:shapiro_r}
\end{equation}
\begin{equation}
s = \sin i
\label{eq:shapiro_s}
\end{equation}
for an orbit inclined at angle $i$ to the observer with a companion of mass $m_\mathrm{c}$. Defining orbital phase $\Phi = 0$ to be at epoch of the ascending node, then the Shapiro delay is greatest at orbital phase $\Phi = 0.25$, when the pulsar is behind the companion, and allows both masses and the orbital inclination to be constrained where it can be measured.

The magnitude of the Shapiro delay is, however, extremely small (sub-$\mu$s) for all cases except where the orbital inclination is very close to $90\degree$. As shown in Table~\ref{fullsolns}, none of the timing residuals for these pulsars are currently below 5~$\mu$s, so the orbital inclination and the constituent masses cannot be constrained.

\begin{figure*}
	\begin{center}
	\includegraphics[width=8.5cm]{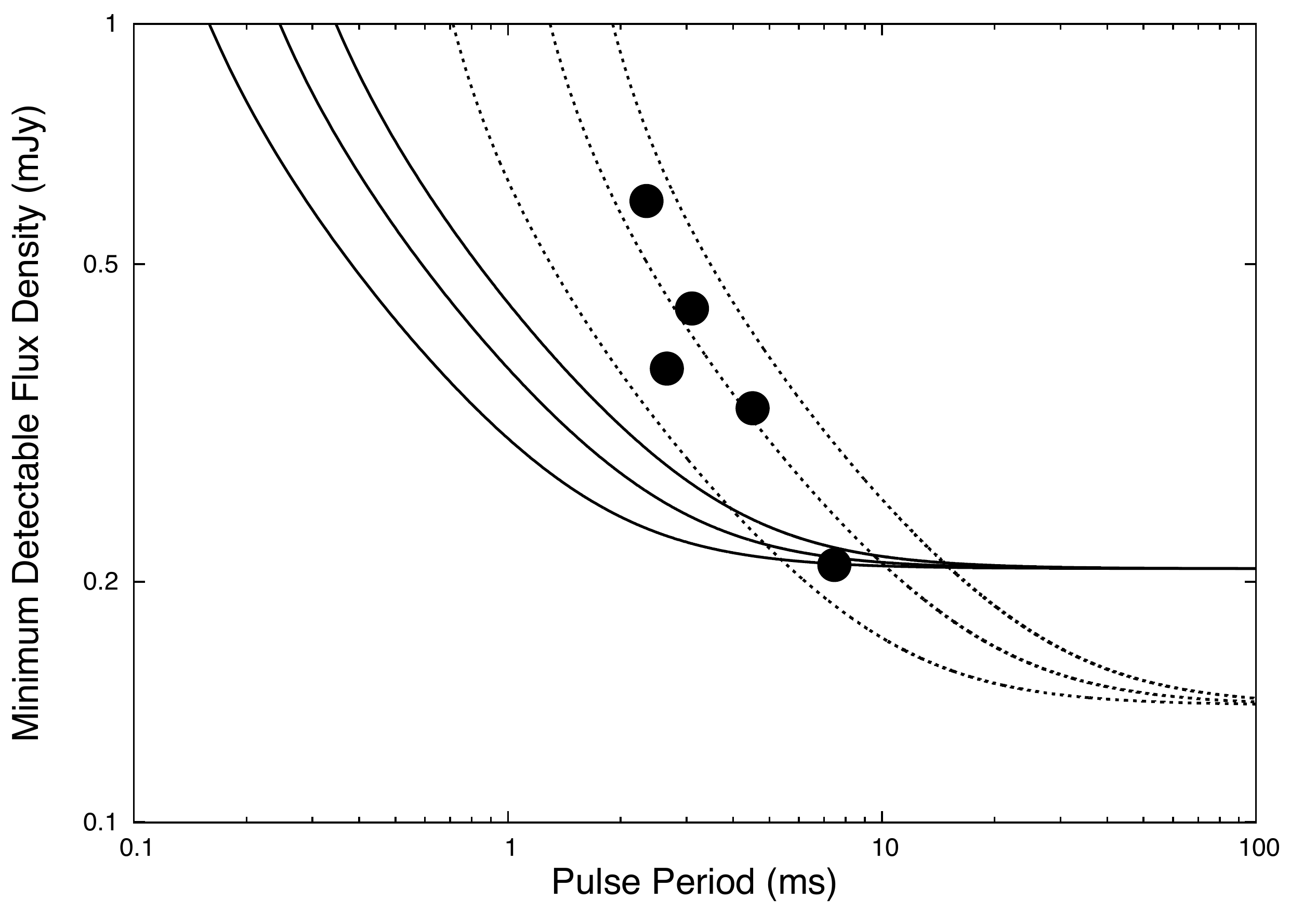}
	\includegraphics[width=8.5cm]{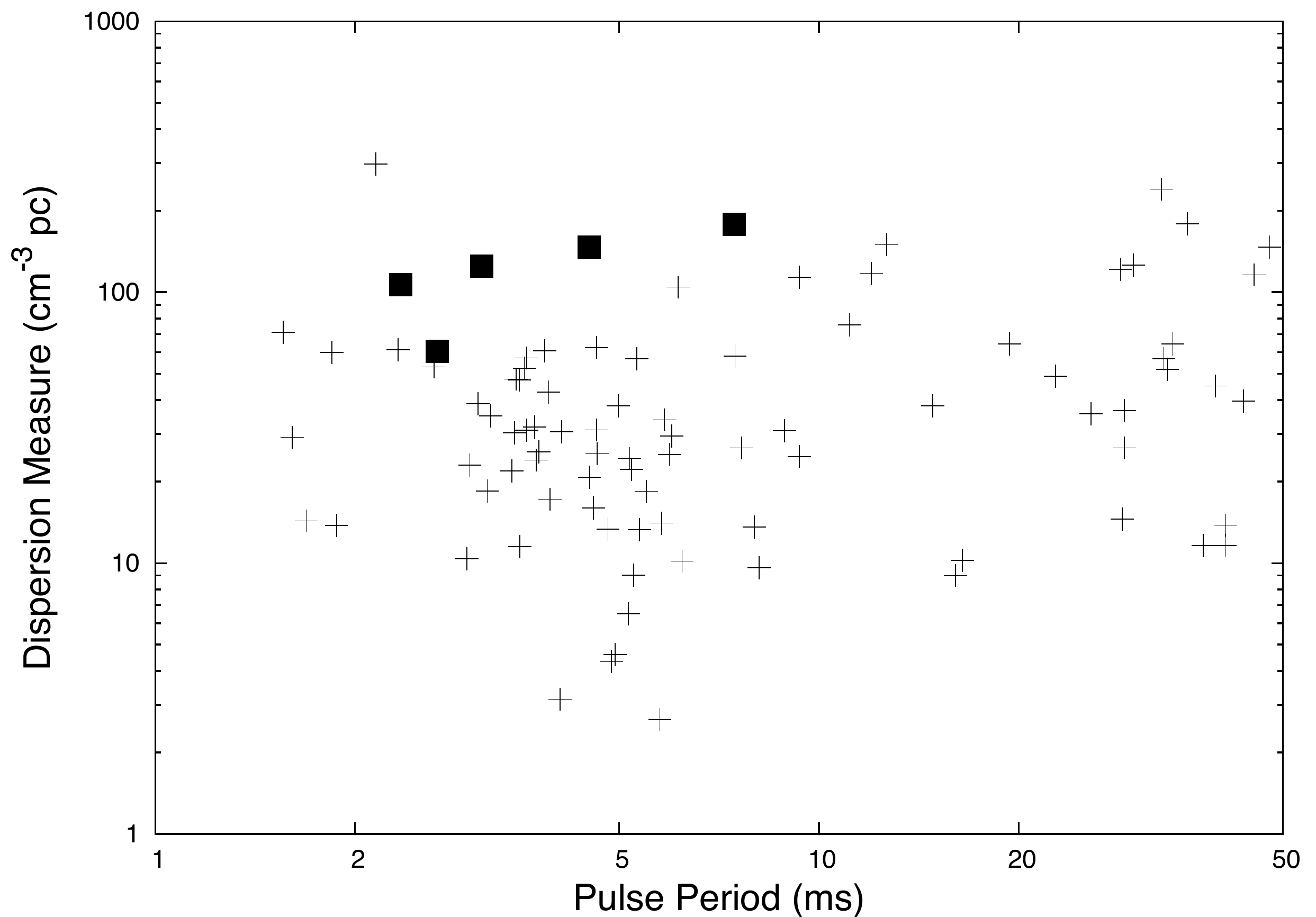}
	\end{center}
	\caption{Left: Minimum detectable flux density, for a detection with S/N = 8, for the HTRU survey (solid line) and PMPS (dotted line) surveys. For each survey, this limiting flux density is plotted for DMs 50, 100 and 150~$\rm{cm}^{-3}\,\rm{pc}$ from left to right, assuming the scattering law of \citet{bcc+04}. The circles indicate the approximate position of the newly-discovered MSPs in this plane. Right: The period and DM of all pulsars with P $<30\,\rm{ms}$, excluding those in globular clusters. Squares indicate the position of our new discoveries in this plane.}
	\label{senspdm}
\end{figure*}

Another GR effect that has been observed for some pulsars is the advance of periastron, $\dot{\omega}$. This is best measured for orbits with a well-defined eccentricity and angle of periastron, $\omega$. Therefore, we only concern ourselves with PSRs~J1125$-$5825 and J1708$-$3506 here. The value of $\dot{\omega}$ \citep{stairs2002} can be calculated, in $\mathrm{rad}\,\mathrm{s}^{-1}$ by
\begin{equation}
\dot{\omega} = 3\left(\frac{G\mathrm{M}_\odot}{c^3}\right)^{^2/_3}\left(\frac{P_\mathrm{b}}{2\pi}\right)^{^{-5}/_3}\frac{1}{1-e^2} \left(m_\mathrm{p} + m_\mathrm{c}\right)^{^2/_3}
\label{eq:omdot}
\end{equation}
for a binary with eccentricity $e$ and orbital period $P_\mathrm{b}$. The values of $\dot{\omega}$ calculated for PSRs~J1125$-$5825 and J1708$-$3506, assuming the median companion mass given in Table~\ref{fullsolns}, are $2 \times 10^{-4}$ and $6 \times 10^{-5}\,\mathrm{\,deg.\,/\,yr}$ respectively. These values compare favourably with the measured $\dot{\omega} = 2.5\times 10^{-4}\,\mathrm{\,deg.\,/\,yr}$ of PSR~J1903+0327 \citep{champion2008}, however, this measurement was obtained with a timing RMS of 1.9~$\mu$s, and for a system with $e=0.44$, which implies that any detection of $\dot{\omega}$ is extremely improbable.


\section{Comparison with discoveries from previous surveys and prospects for multi-wavelength detection}
\subsection{Comparison with previous discoveries}
Following the sensitivity curves presented in \citet{keith2010}, the left-hand panel of Fig.~\ref{senspdm} plots the limiting sensitivity for the HTRU mid-latitude survey as a function of pulse period for pulsars at DMs of 50, 100 and 150~$\mathrm{cm}^{-3}\mathrm{pc}$, and also for the PMPS, responsible for the discovery of a large fraction of the previously-known MSP population. Also plotted are each of the pulsars presented in this paper at their period and flux density. Due to the shorter observation time of the HTRU mid-latitude survey, the limiting flux density at long periods is higher than for the PMPS, however, the HTRU survey is more sensitive to pulsars with shorter rotational periods, due to the short sampling interval of $64\,\mu\rm{s}$ and the high frequency resolution of $0.39~\mathrm{MHz}$ per channel (c.f.\,$250\,\mu\rm{s}$ and $3~\mathrm{MHz}$ for the PMPS). The right-hand panel of Fig.~\ref{senspdm} demonstrates that as well as discovering short-period pulsars, the HTRU survey is sensitive to pulsars with a high DM. Our new discoveries occupy the short-period and high-DM region of the plot, which we can attribute to the short sampling time and narrow frequency channels of our survey system compared with previous blind searches for pulsars \citep[see\,][\,for a full discussion]{keith2010}. Our survey is therefore, on average, finding shorter-period MSPs than previous surveys, but are their companions any different? Recently, \citet{bailes2007} demonstrated that no recycled pulsars existed in which the minimum companion mass exceeds (P/10 ms)~$\mathrm{M}_\odot$. The only MSP to defy this relation was the newly-discovered PSR~J1903+0327 by \citet{champion2008}, which is thought to have had a peculiar history. All of our pulsars are in agreement with this empirical law.

\begin{figure}
	\begin{center}
	\includegraphics[width=8cm]{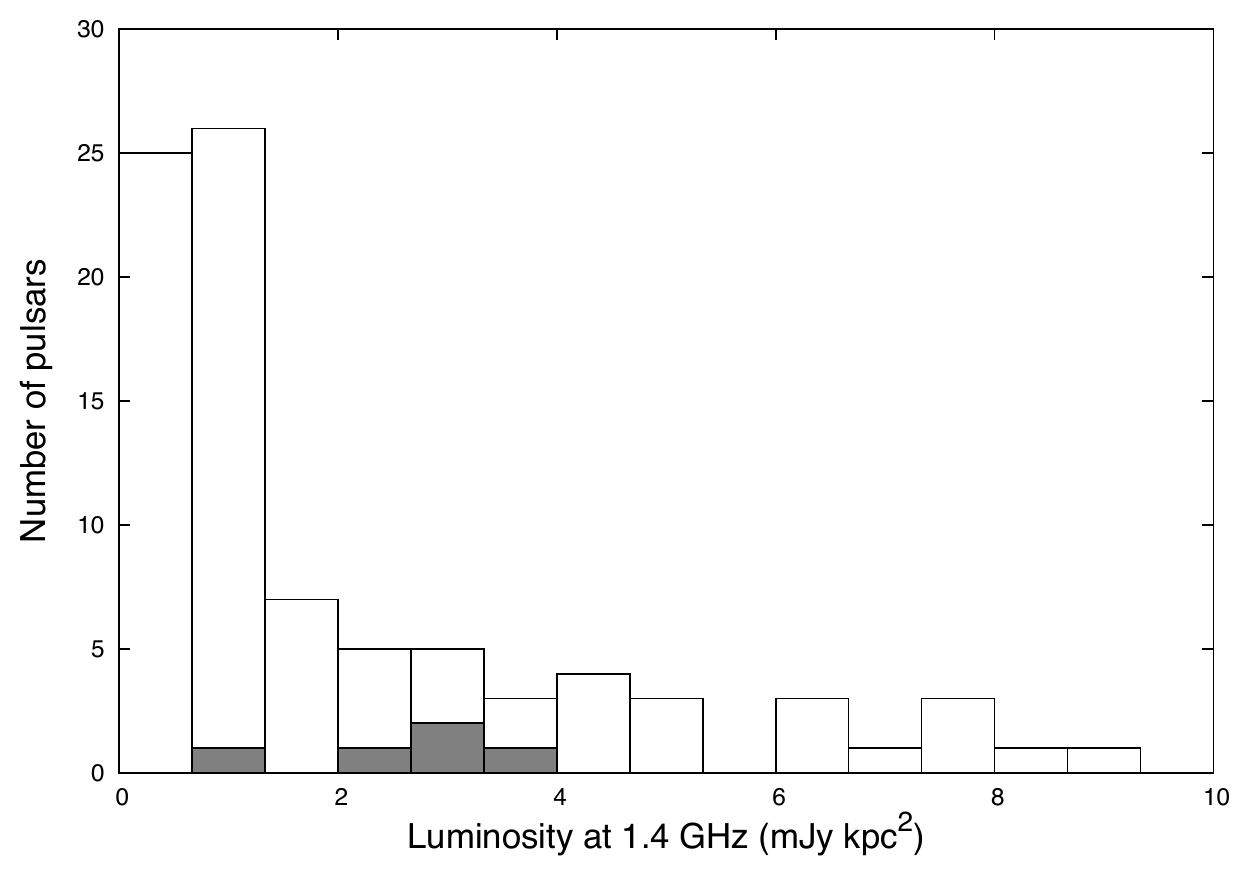}
	\end{center}
	\caption{Histogram of the 1.4~GHz radio luminosity of MSPs ($P \leq 50\,\mathrm{ms}$) in the pulsar catalogue (plain), including those found in globular clusters, and the new discoveries (shaded).}
	\label{luminosity}
\end{figure}

In Fig.~\ref{luminosity} we plot a histogram of the luminosity at 1.4~GHz of the previously-known MSPs (where available, using $P=50\,\mathrm{ms}$ as an upper limit), taken from the pulsar catalogue \citep{mhth05}. A histogram of the luminosities of our discoveries, calculated using distances estimated by the Galactic electron distribution model of \citet{ne2001}, is also plotted with shaded boxes. We have already shown that that the HTRU survey is discovering pulsars of lower flux density, at higher DMs. Since DM is an approximate measure of the distance to a source, this implies that the discoveries have higher luminosities than many of the previously known population.  This is borne out in our results --- of the previously-known MSPs with measured luminosities, $\sim 60\%$ have luminosities below $2\,\mathrm{mJy}\,\mathrm{kpc}^2$. However, four of our newly-discovered MSPs have a luminosity greater than $2\,\mathrm{mJy}\,\mathrm{kpc}^2$. This trend is expected to continue with future discoveries, providing a more complete luminosity distribution in the known population.

\subsection{The possibility of detection with Fermi}
In the last year or so, about 30 MSPs have been discovered with radio follow-up of non-variable previously unidentified point sources from the Fermi Gamma-ray Space Telescope \citep{ray2010}. These discoveries form a complementary group of MSPs to those likely to be found with the HTRU survey, since they are faint, but nearby sources. A check of the Fermi point source catalogue\footnote{http://fermi.gsfc.nasa.gov/ssc/data/access/lat/1yr\_catalog/} indicates that none of these newly-discovered sources have yet been detected as gamma-ray sources, after one year of observations (it is possible that with further observations Fermi may yet make a detection). Compared to pulsars which have been discovered from searches of Fermi point sources \citep{ransom2010}, this lack of detection is consistent with sources which are more distant than the majority of the population. A useful metric with which we can compare the likelihood of a Fermi detection, is by calculating $\sqrt{\dot{E}}/d^2$ \citep[measured in erg kpc$^{-2}$ s$^{-1}$, see\,][\,for details]{fermipsrcat}. For the majority of Fermi-detected MSPs, $\log(\sqrt{\dot{E}}/d^2) \gtrsim 17.5$, and none have been detected where this value is $<16.5$. Of our discoveries, PSR~J1125$-$5825 has the highest value of this metric, with $\log(\sqrt{\dot{E}}/d^2) = 16.6$. Therefore, it is unlikely Fermi will detect this pulsar, or any of our other discoveries, assuming the DM distance holds.

\subsection{Potential for optical detection}
In order to decide whether an optical detection of the companion will be possible for any of the NS-WD systems presented here, we must make use of WD evolutionary models to estimate the luminosity of the companions. Using WD cooling models as described in \citet{fontaine2001}, and using the characteristic age of the pulsar as an indication of the WD age, we are able to estimate the apparent magnitude of the companions, assuming the DM distance given by \citet{ne2001}. These models are highly dependent upon the companion mass, but by using the minimum mass (see Table~\ref{fullsolns}), we are able to put an upper limit on the brightness of the companion. Once we include the effects of interstellar reddening, it becomes unlikely that the companion will be visible, each having an apparent magnitudes greater than $30$. This is a natural consequence of detecting sources at large distances in the Galactic plane.

\section{The Eclipsing Binary MSP PSR J1731$-$1847}\label{sec:1731}
PSR J1731$-$1847 is an eclipsing binary system, and only the fourth to be published, following B1957$+$20, J2051$-$0827 and J1023$+$0038,
to be found outside globular clusters. In order to understand the origin of the eclipsing material, we can approximate the radius of the companion's Roche lobe, $R_\mathrm{L}$, as \citep{egg1983}
\begin{equation}
R_\mathrm{L} = \frac{0.49aq^{2/3}}{0.6q^{2/3}+\ln(1+q^{1/3})} \sim 0.3 \mbox{~R}_\odot
\label{eq:RLeq}\end{equation}
where $q = m_c / m_p$ is the mass ratio of the binary system, and $a$ is the separation of the pulsar from the companion. For this calculation, the companion's mass was assumed to be the minimum possible given the mass function, and the pulsar mass assumed to be 1.4~M$_\odot$. In comparison, the radius of eclipse is approximately 0.9~R$_\odot$, implying that much of the eclipse material is not gravitationally bound to the companion. 

By studying the delay of pulses near to eclipse ingress and egress, we can infer properties of the material causing the eclipses. If we assume the delay, at observing frequency $f$, to be purely dispersive, we can express it as a time delay, $t$, relative to a wave of infinite frequency,
\begin{equation}
t=\frac{e^2}{2\pi m_e c}\frac{\rm{DM}}{f^2}.
\label{eq:disp_delay}\end{equation}
Hence, it is possible to attribute a DM to the delay and, using the orbital separation of the pulsar and its companion, infer the required density of free electrons to cause such a delay, assuming a homogeneous plasma.

\begin{figure}
	\begin{center}
	\includegraphics[width=8cm]{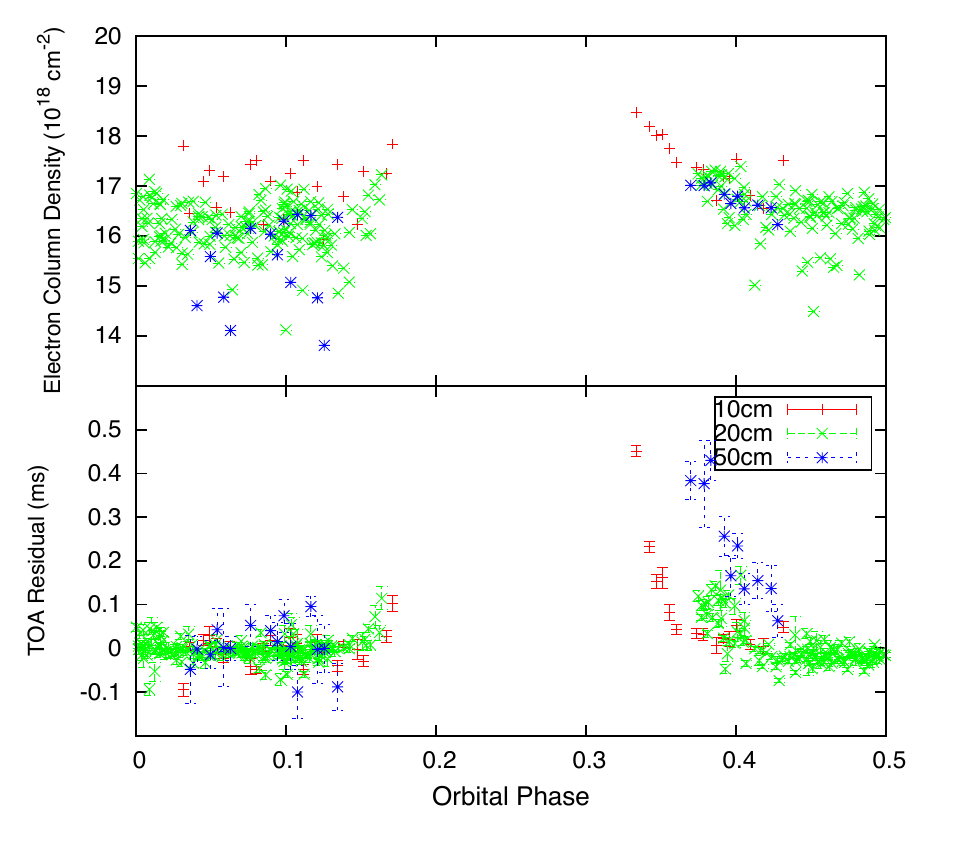}
	\end{center}
	\caption{The bottom panel shows timing residuals for PSR J1731$-$1847 as a function of orbital phase at observing wavelengths of 10 (+), 20 ($\times$) and 50~cm (*). The top panel shows the inferred electron density, from equation (\ref{eq:disp_delay}), for each TOA.}
	\label{1731residuals}
\end{figure}

\subsection{Multi-frequency observations of PSR J1731$-$1847}
In order to probe the eclipse region of this system, observations were performed at wavelengths of 10, 20 and 50~cm using the Parkes radio telescope. In Fig.~\ref{1731residuals} one can see that no pulses are observed between orbital phases of 0.17 and 0.35, while the increased residuals either side of the eclipse show the extra delay to the pulse arrival times, caused by the increased amount of stellar material between ourselves and the pulsar as it approaches superior conjunction, which is usually defined as phase 0.25. At eclipse egress, one can see the additional delay, due to the eclipsing material, reduces most rapidly at 10~cm. Hence, using equation~(\ref{eq:disp_delay}), the added electron density can be computed to be $2.5 \times 10^{17}\,\rm{cm}^{-2}$ at the edge of eclipse. It is also apparent that the ingress and egress are not symmetrical about superior conjunction (orbital phase 0.25), with eclipse egress lasting much longer. This is indicative of the companion's wind being swept behind as it moves along its orbit.

\begin{table*}
	\caption{Eclipse properties for the pulsars PSR J1731$-$1847, PSR J1023$+$0038, PSR J1748$-$2446A, PSR B1957$+$20, and PSR J2051$-$0827.}
		\begin{tabular}{lccccc}
		\toprule
		Parameter & J1731$-$1847 &J1023$+$0038& J1748$-$2446A & B1957$+$20 & J2051$-$0827 \\
		\midrule
		Eclipse Radius ($\mathrm{R}_\odot$) & 0.9 & $\sim0.9$ & 0.8 & 0.75 & 0.3\\
		Roche Lobe Radius ($\mathrm{R}_{\odot}$) & 0.3 & $\sim0.3$& 0.15 & 0.3 & 0.13 \\
		$P_\mathrm{b}$ (d) & 0.311134 & 0.198 & 0.075646 & 0.382 & 0.0991 \\
		$|\dot{P}_\mathrm{b}|$ & $7\times 10^{-11}$ & $2.5\times10^{-10}$ & --- & $3.9 \times 10^{-11}$ & $15.5\times 10^{-12}$ \\
		$|\dot{P}_\mathrm{b}|/P_\mathrm{b}$ ($\rm{s}^{-1}$) & $2.6 \times 10^{-15}$ & $1.5 \times 10^{-14}$ & --- & $1.2 \times 10^{-15}$ & $1.8 \times 10^{-15}$ \\
		$\dot{E}$/$a^2$ ($\rm{erg}\,\mathrm{lt}$-$\rm{s}^{-2}\,\rm{s}^{-1}$)& $1.5\times 10^{33}$ & $3.8\times 10^{33} $ & --- & $2.3\times 10^{33}$ & $4.8\times 10^{32}$\\
		N$_{\rm{e}}$ Maximum ($\rm{cm}^{-2}$) & $3\times 10^{18}$ & $4.6\times 10^{17}$ & $2\times 10^{18}$ & $4\times 10^{17}$ & $4\times 10^{17}$\\
		Anomalous Eclipses & No & Yes & Yes & No & No\\
		\bottomrule
		\end{tabular}
		\label{eclipse_details}
\end{table*}

A number of models have been developed to try to explain the eclipse phenomenon in pulsar systems, many of which are presented in \citet{thompson1994}. The first of these is that the eclipse occurs due to refraction of the pulse emission by the dense electron content in the companion's atmosphere. This model, however, predicts pulses to be heavily delayed (typically by between 10 and 100 ms). We do not observe this amount of delay, and hence can discount the refractive eclipse model for PSR J1731$-$1847. Other eclipse mechanisms proposed by \citeauthor{thompson1994} which we disregard in this work are induced Compton scattering and Raman scattering of the radio beam. When applied to PSR~B1957+20, these models were found to be unviable when explaining eclipses at 1400~MHz, and so we do not explore them further.

Another eclipse mechanism put forward by \citeauthor{thompson1994} is pulse smearing due to the added electron density along the line of sight. The first TOA at eclipse egress is observed at 10~cm, and is delayed by 450~$\mu\rm{s}$. By equation~(\ref{eq:disp_delay}), this implies an added DM of $0.98\,\mathrm{cm}^{-3}\,\mathrm{pc}$, which would cause the pulse to be smeared by 0.77~ms and 1.3~ms at 20 and 50~cm, respectively. This is significantly less than the pulse period of PSR J1731$-$1847, and yet the pulses are not visible; this indicates that pulse smearing is not wholly responsible for the eclipse.

Absorption of the radio emission by the eclipsing material is another potential mechanism for eclipses, via two processes; \begin{inparaenum}[\itshape a\upshape)]
\item cyclotron absorption; and
\item synchrotron absorption.
\end{inparaenum} However, the optical depth of cyclotron absorption is noted by \citeauthor{thompson1994} to have a steep spectral index, and may not be able to explain the total eclipses seen at 10~cm (see Fig.~\ref{1731residuals}). Synchrotron absorption, on the other hand, has a shallower spectral index and may be able to explain these eclipses we observe in PSR~J1731$-$1847 at 10~cm.

\subsection{Comparison with other eclipsing systems}
Including PSR J1731$-$1847 there are now 25 known eclipsing pulsars, many of which are found in globular clusters. Table \ref{eclipse_details} compares the eclipse properties of PSR J1731$-$1847 with those of four of the most well-studied eclipsing binary systems, PSRs J1023$+$0038, B1744$-$24A, B1957$+$20 and J2051$-$0827. All five of these eclipsing systems share similar eclipse and orbital parameters, however, PSRs J1023$+$0038 and J1748$-$2446A display anomalous eclipses indicative of gas outflow from the companion. PSR~J1023$+$0038 displays only a very brief eclipse at $\sim3\,\mathrm{GHz}$, in sharp contrast to the case of PSR J1731$-$1847. However, this is not unreasonable since the electron column density, N$_{\rm{e}}$, in the orbital system of PSR~J1023$+$0038 (calculated from the additional DM given in \cite{asr2009}, shown in Table~\ref{eclipse_details}), is lower by a factor of $\sim 6.5$. Where N$_{\rm{e}} \sim 5\times 10^{17}\,\mathrm{cm}^{-2}$ in the PSR~J1731$-$1847 system, the residuals in the lower panel of Fig.~\ref{1731residuals} shown almost no deviation from zero. It is noteworthy that N$_{\rm{e}}$ takes a similar value for each of these systems, which is not a requirement of any eclipse models.

The comparison of $\dot{E}/a^2$ for these systems shows that the typical energy flux at the companion is around $10^{33}$~$\rm{erg}\,\mathrm{lt}$-$\rm{s}^{-2}\,\rm{s}^{-1}$, more than an order of magnitude higher than the average for all binary systems. In the case that these ``black widow'' systems really are ablating their companions, this shows some evidence for why this process does not occur in all binaries.

PSRs B1957$+$20 and J2051$-$0827 have both been demonstrated to show orbital period variations, with both a $\dot{P}_\mathrm{b}$ and a $\ddot{P}_\mathrm{b}$. In the case of PSR B1957$+$20, these variations take place over a period of years, with $\dot{P}_\mathrm{b} = 1.5\times 10^{-11}$ and $\ddot{P}_\mathrm{b} = 1.4 \times 10^{-18}\,\mathrm{s}^{-1}$  \citep{arzoumanian1994}. PSR J2051$-$0827 has a measured value for the $\dot{P}_\mathrm{b}$ of $-6.4 \times 10^{-12}$, but this value has been observed to fluctuate over the 14 years of observations, however, the variations are not smooth enough to allow measurement of a $\ddot{P}_\mathrm{b}$ (Lazaridis, et al., in prep.). We do not expect, therefore, to be able to measure a $\ddot{P}_\mathrm{b}$ for PSR J1731$-$1847 given that our data span less than a year. However, we have been able to obtain a tentative $\dot{P}_\mathrm{b}$ of $-7(2) \times 10^{-11}$.

The value of $|\dot{P}_\mathrm{b}|/P_\mathrm{b}$ in all these well-studied eclipsing systems are of order $10^{-15}\,\rm{s}^{-1}$. This gives an estimate of the lifetime of these systems of $\sim 10^{7}\,\rm{yr}$, although the variation of $\ddot{P}_\mathrm{b}$ in some of these systems indicates that this lifetime may not be too meaningful.
From Fig.~\ref{fig:ppdot} it can be seen that PSR J1731$-$1847 has one of the highest $\dot{P}$ of the eclipsing and binary pulsars. 

\section{Discussion}
The potential for optical detection of counterparts to these 5 new discoveries is not promising, due to the large distance and the associated Galactic extinction. Prospects for measurement of Shapiro delay are better, although they are dependent upon the orbital inclination. If these systems are highly inclined, PSRs J1125$-$5825 and J1811$-$2405 will have Shapiro delays with the largest amplitude, potentially with as much as 20~$\mu$s delay. In the case of the 6.3 day orbit of PSR J1811$-$2405, this could be detectable, which would allow the measurement of the masses of the pulsar and its companion. For all but the highest inclination angles, however, the delay will not be observable with the current timing precision.

One of these pulsars, PSR J1731$-$1847, is in an eclipsing system with a companion of extremely low mass. Analysis of the eclipsing region shows that the pulse delay, if attributed to dispersion, implies an electron density similar to that measured for PSRs B1744$-$24A and B1957$+$20, which are also in eclipsing systems with low-mass companions.

Two of the pulsars have pulse profiles ideally suited for observations with a pulsar timing array. One of these is, however, PSR~J1731$-$1847, for which there will likely be orbital period changes on timescales of a year or more, and large DM variations that will render it unsuitable for use in pulsar timing arrays. As more observations are made of PSR~J1811$-$2405, it will become clear how well precision timing can be performed with this pulsar, although the current RMS is not small enough to merit inclusion in a timing array. The discovery of distant and bright MSPs is also important for a rounded view of the MSP population and their distribution in the Galaxy. This has implications for searches with future telescopes such as the SKA \citep{smits2008}, and the pulsar timing arrays with these future facilities.

These discoveries have been made with only $\sim 30$\% of the HTRU mid-latitude survey data searched for pulsars, indicating that we should expect to find more such objects in our data. This compares favourably to simulations presented in \citet{keith2010}, which indicated we should expect a total haul of $\sim 30$ new MSPs in the mid-latitude survey.

\section{Conclusions}
We present the discovery of 5 MSPs in the HTRU survey. These MSPs have short periods and DMs among the highest in the known population. One of the pulsars, PSR~J1731$-$1847, displays regular eclipses in its 0.3~d orbit and with a companion of minimum mass $0.04\,\mathrm{M}_\odot$, appears to be a member of the `black widow' group of MSPs.

The sensitivity to such objects can be attributed to the fast sampling rate and narrow filterbank channels of the hardware used for the HTRU survey. 

\section*{ACKNOWLEDGEMENTS}
The Parkes Observatory is part of the Australia Telescope which is funded by the Commonwealth of Australia for operation as a National Facility managed by CSIRO. SDB gratefully acknowledges the support of STFC in his PhD studentship. We thank Gemma Janssen for generously giving up some observing time and performing observations with the Westerbork telescope, and Cees Bassa for useful discussions regarding the white dwarf companions.
\newpage

\bibliography{msps2010}
\bibliographystyle{mnras}

\label{lastpage}

\end{document}